\documentclass[twocolumn]{aastex631}
\usepackage{amsmath}

\shorttitle{Star-disk interaction and kinematics of S-stars}
\shortauthors{Fan et al.}

\begin{document}

\title{The Interaction Between Stars and Past AGN Disk: Possible Explanation for the Kinematic Distributions of S-stars in the Galactic Center}

\author[0000-0001-7350-8380]{Xiao Fan$^{\dagger}$}
\email{$\dagger$: hustfan@hust.edu.cn}
\author[0000-0003-4773-4987]{Qingwen Wu$^*$}
\email{* Corresponding author: qwwu@hust.edu.cn}
\author[0000-0002-2581-8154]{Jiancheng Wu}
\author[0009-0003-0516-5074]{Xiangli Lei}
\author[0000-0001-5019-4729]{Mengye Wang}
\affiliation{Department of Astronomy, School of physics, Huazhong University of Science and Technology, Luoyu Road 1037, Wuhan, China}
\author[0000-0002-5323-2302]{Fulin Li}
\affiliation{Purple Mountain Observatory, Chinese Academy of Sciences, Nanjing 210023, China} 
\affiliation{University of Science and Technology of China, Hefei 230026, China}

\begin{abstract}
The presence of young stars, aged around several million years and situated within the range of $\sim 0.04-1$ pc from our Galactic center raises a question about their origins and dynamical evolutions. Their kinematics provide an opportunity to explore their formation or possible subsequent dynamical evolution. If Sagittarius A* was active in the past as suggested by several observations, the accretion disk may have a significant impact on the dynamics of stars in the Galactic center. The drag force exerted on stars during star–disk interaction could lead some of them to sink into the accretion disk, and these embedded stars will rapidly migrate inward and eventually be disrupted within $\sim10^5$ yr. This could roughly explain the absence of stars within $2.5 \times 10^4 R_{\rm g}$ ($\sim$ 1000 au). Additionally, Kozai–Lidov oscillations, induced by the gravitational perturbation of the disk, could contribute to the bimodal distribution of S–star inclinations and drive a majority of stars into high eccentricity orbits.

\end{abstract}

\keywords{Active galactic nuclei; Galactic center; Stellar dynamics}

\section{Introduction}
It is widely believed that the supermassive black holes (SMBHs) reside at the center of most galaxies \citep[e.g.,][]{ARAAHo2013}. These SMBHs grow through mergers or gas accretions, occasionally shining as active galactic nuclei (AGNs) during the accretion phase \citep[e.g.,][]{Soltan1982}. Even though most SMBHs or galaxies are quiescent, several empirical correlations have been found between the SMBHs and galaxy bulges, which support a strong connection and interplay between the growth of SMBHs and galaxy bulges \citep[e.g.,][]{Magorrian1998,Ferrarese2000,Gultekin2009}. At the center of our Galaxy lies an SMBH (known as Sagittarius A*, Sgr A*) with mass around $ 4 \times 10^6 \, {M}_{\odot}$ \citep[e.g.,][]{Genzel1997,Ghez1998,Gillessen2017}, and the current bolometric luminosity is quite low \citep[$L_{\rm bol}\sim 10^{-9}L_{\rm Edd}$, where $L_{\rm Edd}$ is Eddington luminosity; e.g.,][]{Melia2001,Broderick2009}. Although Sgr A* is quiescent currently, several observations suggest that Sgr A* may be much brighter in the past. X–ray reflections from dense gas in the Galactic center suggest a past short-lived flaring activity on timescales of hundreds or thousands of years \citep{Sunyaev1993,Ponti2010,2018A&A...610A..34Chuard,Marin2023}. Some evidence also reveals significant activities of Sgr A* on much longer timescales. \citet{Bland2013} proposed that the ${\rm H \alpha}$ emission along the Magellanic stream was triggered by a ``Seyfert–like flare" from Sgr A* 1-3 Myr ago. Furthermore, two giant gamma–ray bubbles were discovered above and below the Galactic center in 2010 through analyzes of Fermi Large Area Telescope data, which are called the ''Fermi Bubbles" \citep{Su2010}. Although the physical origin of the Fermi Bubbles remains uncertain, their features are consistent with being inflated several million years ago by either an AGN jet \citep[e.g.,][]{Guo2012} or a wind–driven outflow \citep[e.g.,][]{Zubovas2012,Mou2014}.

Sgr A* is surrounded by a dense cluster of young stars in the central parsec \citep[e.g.,][]{Genzel1997,Ghez1998,Gillessen2017}. One population is luminous massive OB/Wolf–Rayet(W–R) stars locating at $\sim \, 0.04- 0.5$ pc \citep{Eckart1999,Paumard2006,Bartko2009,Yelda2014,Von2022}. These young stars \citep[e.g., 6$\pm$2 Myr,][]{Paumard2006} have eccentricities from 0 to 0.7 and appear to reside in a possible warped disk with clockwise rotation (so-called the clockwise stellar disk, CWD), which was first discovered by \citet{levin2003}. Interior to the young star disk lies another more conspicuous population of B stars with highly eccentric orbits in the inner $1{''}$  \citep[$\sim 0.04 \, \rm{pc}$;][]{genzel2003,ghez2003,Gillessen2009,Gillessen2017}, which are often called as ``S-stars" \citep{Eckart1996}. Using the results of high-resolution spectroscopy, \citet{Habibi2017} found that the mass of S-stars is about 8-14 ${M}_{\odot}$ and the age is about 6.6 Myr for the star S2, which is compatible with the age of stars in the CWD. For other S–stars, ages are normally less than 15 Myr.

The presence of young S-stars in the vicinity of Sgr A* is a challenge for traditional star formation theories \citep[e.g., ``Paradox of the Youth";][]{ghez2003}. Due to tidal forces exerted by the central SMBH, the typical molecular clouds within the central parsec is hard to collapse to form stars \citep[e.g.,][]{Sanders1992,Morris1993}. One possible origin of S-stars is that they are remnants of disrupted stellar binaries, which is supported by the observations of hypervelocity stars possibly ejected from the Galactic center \citep{Hills1988,Gould2003,Lockmann2008,madigan2009,Koposov2020}. This scenario predicts very high eccentricities, and some secular processes, such as scalar resonant relaxation or the perturbation from a cluster of stars or stellar–mass black holes, could thermalize the eccentricities \citep{Perets2009,generozov2020}. 


It has been suggested that the outer regions of accretion disks are gravitationally unstable and could trigger fast star formation \citep[e.g.,][]{paczynski1978,Kolykhalov1980,Chen2023}. The broad emission lines of AGNs indicate that the vicinities of SMBHs are metal rich (e.g., several times the solar abundances) and seem to be independent of redshifts \citep{Hamann1992,HF1999,nagao2006,xufei2018,wangshu2022}, which is widely explained by rapid and intense star formation and evolution in AGN disk \citep{Artymowicz1993,wjm2010,wjm2011,wjm2012,wjm2023,Fan2023}. \citet{Paumard2006} suggested that in situ star formation in the accretion disk is in good agreement with the disk structure and stellar features of the stars in the CWD \citep[see also][]{levin2003,Nayakshin2007,Bonnell2008,mapelli2012}. \citet{levin2007} proposed that S–stars were born in the disk and then migrated inward, and their orbits have been randomized by resonant relaxation.

The origin and evolution of young stars in the Galactic center are quite puzzling. \citet{Ali2020} presented a detailed analysis of the kinematics of 39 stars (mostly S-stars including seven stars belonging to the CWD) surrounding the Sgr A*, and they found that the distribution of the inclinations significantly deviates from an isotropic distribution. They argued that S–stars are arranged into two almost orthogonal disks. If Sgr A* has experienced the ``Seyfert–like" phase millions of years ago, the once existed accretion disk will influence the stellar evolution and the dynamics of stars significantly. 
The main purpose of this paper is to explore the possible causes of the observational features of S–stars in the Galactic center via studying the dynamics of a spherical star cluster interacting with the possibly once existed accretion disk. In Section \ref{sec:2}, we present the accretion disk model and secular dynamics of stars affected by the star-disk interaction. In Section \ref{sec:3}, we simulate the orbital evolution of stars interacting with the accretion disk and compare the results with the observations of S–stars. The conclusion and discussion are presented in Section \ref{sec:4}.

\section{The models for star-disk interaction}\label{sec:2}
In the AGN phase, accretion disks surrounding SMBHs are much hotter and denser than most components of the interstellar medium. The accretion disk may play a crucial role in both the evolution and dynamics of stars in the central parsec of Sgr A*. Stars with nonzero inclinations relative to the disk, which we call off–disk stars, will periodically cross the accretion disk. The drag force exerted on stars during each crossing will attenuate the semimajor axis, eccentricities, and inclinations of stars \citep[e.g.,][]{Macleod2020,Fabj2020,Nasim2023,generozov2023}. Meanwhile, the AGN disk itself, serving as a gravitational source, will perturb the orbits of off-disk stars and lead to a dynamical behavior similar to Kozai–Lidov (KL) oscillations \citep{kozai1962,Lidov1962,Subr2005,Terquem2010,Liu2020}, but which may be suppressed by the periapsis precession due to general relativity (GR) and the gravity of the stellar bulge \citep{Ford2000,Fabrycky2007,Chang2009,Naoz2013b}. Unlike the off-disk stars, the dynamics of stars embedded in the disk are dominated by the effect of migration, i.e., stars migrate inward due to the excitation of spiral density waves. It should be noted that the migration is only efficient for stars embedded in the disk, and the migration timescale is \citep{Tanaka2002}
\begin{equation}
\tau_{\rm mig}= \Omega_{*}^{-1}H^2\left ( \frac{m_*\Sigma_{\rm d} a_*^4}{M_{\rm BH}^2}  \right )^{-1},
\end{equation}
where $M_{\rm BH}$ is the mass of the central black hole, $a_*$, $\Omega_{*}$, and $m_{*}$ are the orbital semimajor axis, the Keplerian angular velocity, and the mass of the star respectively. $H$ and $\Sigma_{\rm d}$ are the half height and surface density of the disk. For stars with nonzero inclinations or eccentricities, the effect of migration is negligible \citep[e.g.,][]{Rein2012,Fendyke2014,Arzamasskiy2018}. In this paper, we focus on the secular dynamics of a star, considering the drag force on stars as they cross the disk, the gravitational perturbation from the disk, and the periapsis precession due to GR and stellar bulge.

\subsection{Accretion disk model}
We adopt the steady-state accretion disk model that extends to the outer gravitationally unstable region as proposed by \citet{sirko2003}. The inner part of the disk is the classical standard $\alpha$-disk \citep{shakura} and the outer gravitationally unstable disk maintains $Q\equiv c_{\rm{s}}\Omega_{\rm d} /\pi G \Sigma_{\rm d}=1$ \citep{1964ApJ...139.1217Toomre} due to the possible auxiliary heating mechanisms \citep[e.g., the energy released by stars; see also][]{thompson2005}, where $c_{\rm s}$ is the sound velocity, $\Omega_{\rm d}$ and $\Sigma_{\rm d}$ are Keplerian angular velocity and surface density of the disk respectively. The accretion rate is $\dot{M}_{\rm BH} =f_{\rm Edd} \dot{M}_{\rm Edd}$, where $f_{\rm Edd}$ is dimensionless accretion rate and $\dot{M}_{\rm Edd}$ is the Eddington accretion rate. In this work, we adopt $M_{\rm BH}=4\times 10^6 \, {M}_{\odot}$ and viscosity parameter $\alpha_{\rm vis}=0.1$ as the typical values. For the opacity of the disk, we use the opacity tables of \citet{iglesias1996} for high temperatures and those of \citet{Ferguson05} for low temperatures.

\subsection{Drag force}
The star will suffer a drag force when crossing the accretion disk due to the relative velocity between the star and the gas of the accretion disk. In the case of less–massive stars, the gas collides directly with stars to generate an aerodynamic drag \citep{Adachi1976},
\begin{equation}
F_{\rm aero}=\frac{1}{2}C_D \pi r_{*}^2 \rho_{\rm d} v_{\rm rel}^2,
\end{equation}
where $r_{*} $ is the stellar radius, $v_{\rm vel}$ is the relative velocity of the star to the accretion disk, $C_D$ is the nondimensional drag coefficient, and $\rho_{\rm d} \equiv \Sigma_{\rm d}/2H$ is the density of the accretion disk. The parameter $C_D=2$ is adopted since the relative velocities of stars are supersonic in most cases \citep{Adachi1976}. For more massive stars, the gas in the disk will be gravitationally scattered by the star, causing the star to suffer dynamical friction \citep{Ostriker1999,Edgar2004}
\begin{equation}
F_{\rm dyn}=\frac{4\pi G^2m_*^2 \rho_{\rm d}}{v_{\rm rel}^2}.
\end{equation}

The above two types of drag force can be written together as 
\begin{equation}
F_{\rm drag}=\pi r_{\rm drag}^2 \rho_{\rm d} v_{\rm rel}^2,
\end{equation}
where $r_{\rm drag}= {\rm max}(r_{*},{2Gm_{*}}/{v_{\rm rel}^2})$ is the actual interactional radius of stars with the accretion disk \citep[see also][]{Macleod2020,generozov2023}.

In one orbital period, the star crosses the accretion disk twice. The distances of crossing points from the SMBH are $r_{\pm}={a_*(1-e_*^2)}/({1\pm e_*{\rm cos}\, \omega_*})$, where $e_*$ is the eccentricity of the star and $\omega_*$ is the argument of periapsis of the star. The true anomaly of the star at the closer and farther crossing points are $\psi_+=- \omega_*$ and $\psi_-=- \omega_* + \pi$ respectively. The subscript $+ (-)$ denotes the closer (farther) crossing point. Therefore, the orbital velocity of stars at the cross points is $v_{\pm}=\sqrt{GM_{\rm BH}\left( {2}/{r_{\pm}} -{1}/{a_*}\right)}$. \cite{Adachi1976} presented a detailed derivation of the gas drag effect of an object immersed in the gas cloud; following their works, the relative velocities between stars and the disk at the crossing points are
\begin{equation}
\begin{split}
v_{\rm vel, \pm}= &\frac{\sqrt{G M_{\rm BH}/a_*}}{(1-e_*^2)^{1/2}} [2\pm3e_*{\rm cos}\, \omega_* \\& +e_*^2 -2(1\pm e_*{\rm cos}\, \omega_*)^{3/2}{\rm cos}\, i_*)]^{1/2},
\end{split}
\end{equation}
where $i_*$ is the inclination of the stellar orbits relative to the plane of the accretion disk.

The drag force exerted on the star by the accretion disk will cause the inclination, eccentricity, and semimajor axis of the star to decay. The decays of the orbital parameters at the two crossing points are respectively given by
\begin{equation}
\begin{split}
{da}_{\rm *, \pm}&= -\frac{a_*}{\tau_{\rm  \pm}}\frac{2}{1-e_*^2}[1\pm2 e_* {\rm cos}\, \omega_*\\&+e_*^2-(1\pm e_* {\rm cos}\, \omega_*)^{3/2}{\rm cos}\, i_*]  t_{\rm cross, \pm}   ,
\end{split}
\end{equation}

\begin{equation}
\begin{split}
{de}_{\rm *, \pm}&= -\frac{1}{\tau_{\rm \pm}}[ 2e_* \pm2 {\rm cos}\, \omega_* \\&-\frac{e+e{\rm cos}^2\, \omega_* \pm 2{\rm cos}\, \omega_*}{(1\pm e_*{\rm cos}\, \omega_*)^{1/2}}  {\rm cos}\, i_*]t_{\rm cross, \pm}     ,
\end{split}
\end{equation}
and
\begin{equation}
\begin{split}
{di}_{\rm *, \pm}= -\frac{1}{\tau_{\rm \pm}} \frac{{\rm sin}\, i_* }{(1\pm e_*{\rm cos}\, \omega_*)^{1/2}} t_{\rm cross, \pm}  ,
\end{split}
\end{equation}
where 
\begin{equation}
\tau_{\pm} = \frac{m_*}{ \pi r_{\rm drag}^2 \rho_{\rm d} v_{\rm rel, \pm}},
\end{equation}
and $t_{ \rm cross, \pm}=2H/v_{\pm}{\rm sin} \, i_*$ is the time for the star to cross the AGN disk. Considering the orbital decays at closer and father crossing points, the damping timescale due to drag force can be written as
\begin{equation}\label{eq:tau_drag}
\frac{1}{\tau_{\rm drag}} \simeq \frac{t_{ \rm cross, +}}{\tau_{+}t_{\rm orb}}+\frac{t_{ \rm cross, -}}{\tau_{-}t_{\rm orb}},
\end{equation}
where $t_{\rm orb}=2\pi \sqrt{a_*^3/G M_{\rm BH}}$ is the orbital period of the star. The star will sink into the disk or fall into the black hole due to the drag force if the interaction timescale is longer than $\tau_{\rm drag}$.

\subsection{Kozai–Lidov oscillation} 
In a hierarchical three body system, the inner object orbiting the massive central body is perturbed by a distant tertiary object, the inner object will periodically exchange its eccentricity and orbital inclination to ensure the conservation of $z$–component of the angular momentum. This dynamical phenomenon is known as the Kozai–Lidov mechanism \citep{kozai1962,Lidov1962}, occurring on a timescale much longer than the orbital period. 
In the case that the gravitational perturbation is from a disk and the size of the disk is several times larger than the orbit of inner object, the secular dynamics of the inner object are identical to the classical KL effect \citep[e.g.,][]{Subr2005,Terquem2010,Liu2020}.

\citet{naoz2013} suggested that in the case of a perturber in a circular orbit, such as a disk, it is sufficient to describe the motion of the inner object by considering only the quadrupole-level terms of the Hamiltonian. In order to derive the equations of motion for a star perturbed by a disk, the accretion disk can be viewed as a superposition of a sequence of disk rings and thus the perturbation exerted on a star can be regarded as the combined effects of each disk ring \citep[][]{Chang2009,Chen2014}. Following the work of \citet{naoz2013}, the evolution equations of the eccentricity, inclination, and argument of periapsis of the star are
\begin{equation}\label{eq:edot}
\begin{split}
\dot{e}_{\rm *,KL}= 30 \cdot C_{\rm KL} \, {\rm sin}^2 \, i_* \,  {\rm sin} (2\omega_*) \, e_* \sqrt{1-e_*^2},
\end{split}
\end{equation}
\begin{equation}
\begin{split}
\dot{i}_{\rm *,KL}= -30 \cdot C_{\rm KL} \, \frac{{\rm sin}^2 \, i_* \, {\rm sin}(2\omega_*)}{{\rm tan}\, i_*} \frac{e_*^2}{ \sqrt{1-e_*^2}},
\end{split}
\end{equation}
and
\begin{equation}
\begin{split}
\dot{\omega}_{\rm *,KL} &=6\cdot C_{\rm KL} \frac{1}{ \sqrt{1-e_*^2}}  \times \biggl[ 4 \, {\rm cos} ^2i_*\\& +
(5 \, {\rm cos}(2\omega_*)-1) (1-e_*^2-{\rm cos}^2 \, i_*) \biggr]  ,
\end{split}   
\end{equation}
where the coefficient
\begin{equation}
C_{\rm KL}=\int_{2a_*}^{R_{\rm out}} \frac{1}{16} \sqrt \frac{G \, {\delta m}^2}{{M_{\rm BH}}} \frac{a_*^{3/2}}{r^3}dr,
\end{equation}
which takes into account the combined effect of each disk ring; $R_{\rm out}$ is the outer boundary of the accretion disk, and $\delta m =2 \pi r \Sigma _{\rm d} dr $ is the mass of the disk ring located at radius, $r$, from central SMBH. The equations above have been simplified under the condition $m_*  \ll \delta m \ll M_{\rm BH}$. It should be noted that the perturbations from disk rings inside the orbits of stars do not significantly affect the stellar orbit \citep[see Eq.15 in][]{naoz2017}; therefore, we adopt $2a_*$ as the lower limit of the integral to ensure that only the effects of rings completely outside the orbits of stars are taken into account. \citet{Antognini2015} presented the period of KL oscillations is
\begin{equation} \label{KL timescale}
\tau_{\mathrm{KL}} \simeq \frac{1}{15C_{\rm KL}}.
\end{equation}

\subsection{Precession due to GR and stellar bulge}
The effect of GR and the gravity of the stellar bulge will lead to periapsis precession of the star. In the Galactic center, the radial mass distribution of the stellar bulge follows a power–law distribution with index $-\gamma$, $\rho_{\rm *}(r)=\rho_0 (r/r_0)^{-\gamma}$. Based on the best fitting of observations of stellar surface density distribution, \citet{Genzel2010} found that $\rho_0=1.35\times 10^6 \, {M}_{\odot} {\rm pc}^{-3}$, $r_0=0.25 \,{\rm pc}$, and $\gamma=1.3$. Therefore, the periapsis precession of the star can be written as
\begin{equation}
\begin{split}
\dot{\omega}_{\rm *,prec}= \frac{3(GM_{\rm BH})^{3/2}}{a_*^{5/2}c^2(1-e_*^2)}-\kappa \frac{M_{\rm bulge}}{M_{\rm BH}}\sqrt[]{\frac{GM_{\rm BH}}{a_*^3} } ,    
\end{split}
\end{equation}
where the first term is the GR precession term \citep{Fabrycky2007,Naoz2013b} and the second term is the stellar bulge precession term \citep{Ivanov2005}. $M_{\rm bulge}=\int_{0}^{a_*}  4 \pi r^2 \rho_*(r) dr$ is the mass of the bulge inside the star, $\kappa=\Gamma(5/2-\gamma)/(\sqrt{\pi}\Gamma(3-\gamma))$ and $\Gamma$ is the gamma function. Note that the precession due to GR and stellar bulge takes place on opposite direction, and for stars with small orbits, the precession is dominated by GR effects, whereas for stars with large orbits, the precession is dominated by the effects of stellar bulge. The precession timescale is given by
\begin{equation}
\tau_{\rm prec} \simeq |1/\dot{\omega}_{\rm *,prec}| ,
\end{equation}
the KL oscillations will be suppressed if this timescale is several orders of magnitude shorter than the timescale of KL oscillations \citep[e.g.,][]{Ford2000,Fabrycky2007,Chang2009,Naoz2013b}.

\begin{figure}
\centering
\includegraphics[scale=0.35]{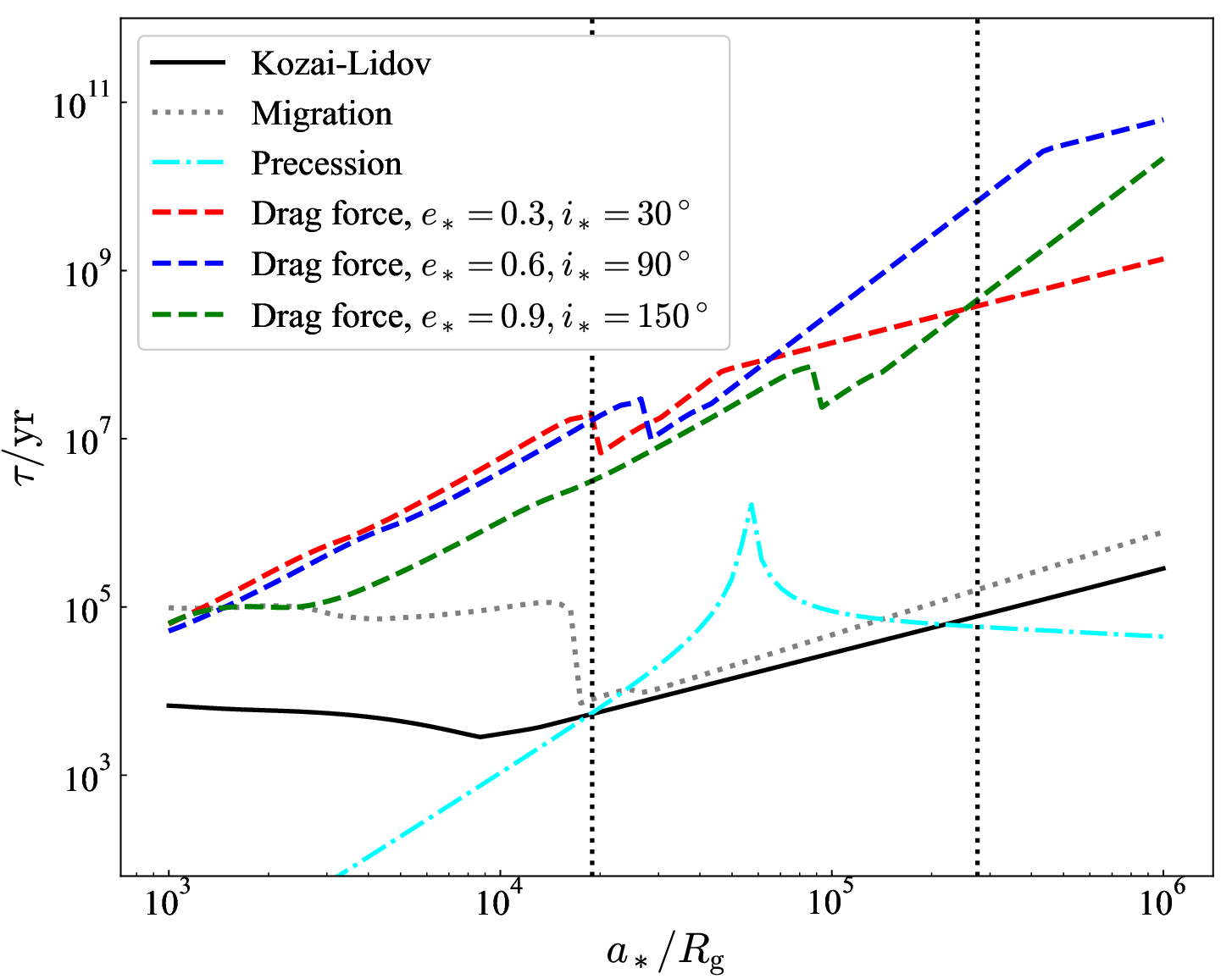}
\caption{Timescales for different effects on stars with different orbital semimajor axis. The black solid line shows the timescale of the KL oscillations due to the perturbation of the disk. The dashed lines represent the damping timescale of stars due to drag force for the cases $e_*=0.3,i_*=30^\circ$ (red), $e_*=0.6,i_*=90^\circ$ (blue), and $e_*=0.9,i_*=150^\circ$ (green). The gray dotted line illustrates the migration timescale for stars embedded in the accretion disk. The cyan dash–dotted line indicates the timescale of periapsis precession for stars with typical value of $e_*=0.7$. Between the two vertical dotted lines is the range of S-stars. In the calculations, the typical values of $f_{\rm Edd}=0.1$, $\omega_*=\pi/4$, and $m_*=10 \, {M}_{\odot}$ are adopted.}
\label{fig:tau_compare}
\end{figure}

\section{Results}\label{sec:3} 
\begin{figure*}
\includegraphics[scale=0.47]{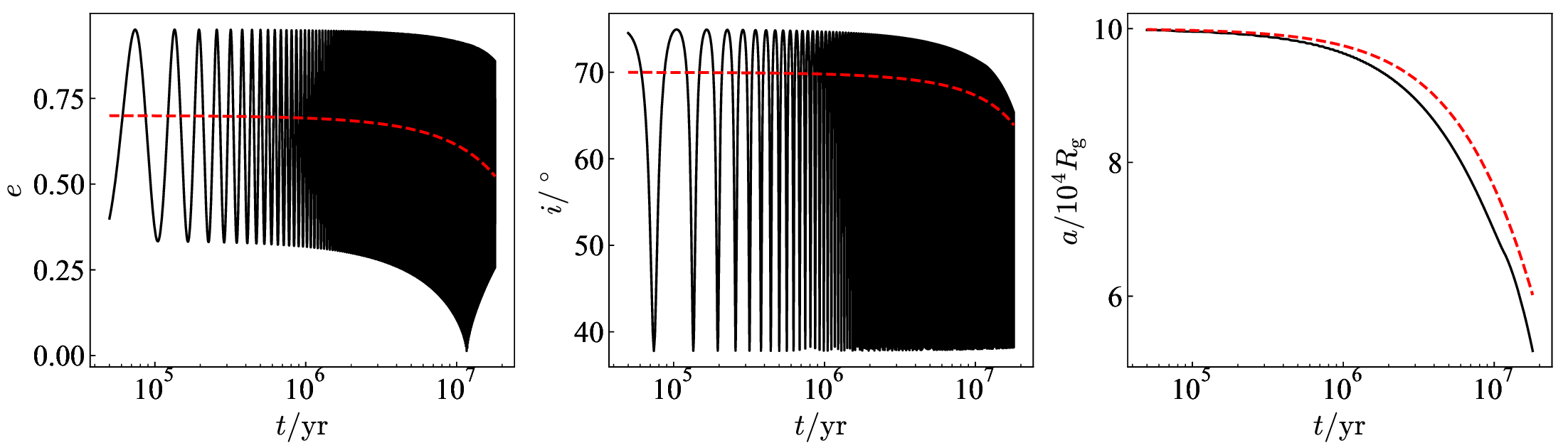}
\caption{The evolution of a star with initial orbital parameters $a_{*}=10^5 \, R_{\rm g}$, $e_{*}=0.7$, and $i_{*}=70^\circ$. The black solid line takes into account all the effects and the red dashed line takes into account only the effects of the drag force.}
\label{fig:evolution}
\end{figure*}
In Figure \ref{fig:tau_compare}, we present the timescales of several effects on stars considered in this paper. It can be found that the migration timescale for stars embedded in the accretion disk is $\sim10^4-10^{5}$ yr, which is much shorter compared to the age of S-stars. Therefore, the stars in the disk will quickly migrate inward and finally be tidally disrupted or swallowed by the SMBH \citep[e.g.,][]{wangmengye2023}. The drop of migration timescale at $\sim 10^4 R_{\rm g}$ is caused by the abrupt changes in disk density that caused by hydrogen ionization instability. For those more general off-disk stars, the effect of migration is negligible, and the dynamics of stars are governed by the combination of KL oscillations and drag force due to crossing the disk, as well as the periapsis precession due to GR and stellar bulge. The KL timescales are $\sim10^4-10^5$ yr for $a_*\sim10^3-10^6 \, R_{\rm g}$, $R_{\rm g}=GM_{\rm BH}/c^2$ is the gravitational radius. In calculating the KL timescale, $R_{\rm out} = 10^7 \, R_{\rm g}$ is taken as the outer boundary of the accretion disk. In fact, the outer boundary does not significantly affect this timescale since the effects of the part of the disk far from the star are negligible (see Equation \ref{KL timescale}). To ensure the occurrence of the KL oscillations, it is sufficient for the outer radius of the accretion disk to be several times larger than the semimajor axis of the star. Between the two vertical dotted lines is the range of S-stars, where the precession timescale is longer than the KL timescale. The precession timescale reaches its maximum at $a_* \sim 6\times 10^4 \, R_{\rm g}$, where the precession induced by stellar bulge is counteracted by the effect of GR. For those stars with larger or smaller orbits (e.g., stars beyond the vertical lines), the precession will be dominated by stellar bulge or GR, respectively. When the precession timescale is shorter than the KL timescale, the KL oscillations will be suppressed due to the periapsis precession \citep[e.g.,][]{Ford2000,Fabrycky2007,Chang2009,Naoz2013b}. The damping timescales for the effect of drag are $\sim10^5-10^{10}$ yr for $a_*\sim10^3-10^6 \, R_{\rm g}$, which is much longer than the KL timescale. If the interaction time between a star and an accretion disk is long enough, the star will sink into the disk and migrate inwards, eventually be tidally disrupted or swallowed by the central SMBH.

An example of the evolution of a star is shown in Figure \ref{fig:evolution}. It can be found that the star undergoes oscillations during the decay of eccentricity, inclination, and semimajor axis. As the semimajor axis decays, the precession induced by the stellar bulge gradually weakens and is counteracted by the effects of GR. Therefore, the amplitude of the KL oscillations increases progressively and reaches its maximum around $ 10^7$ yr. Subsequently, with further orbital decays, GR dominates the precession, leading to the resumption of the suppression of KL oscillations. In contrast to orbital evolution considering only the drag force (illustrated by the red dashed line), additionally taking into account of KL mechanism leads to more–rapid orbital decays, since the KL mechanism can pump stars into more–eccentric orbits.

\begin{figure*}
\centering
\includegraphics[scale=0.47]{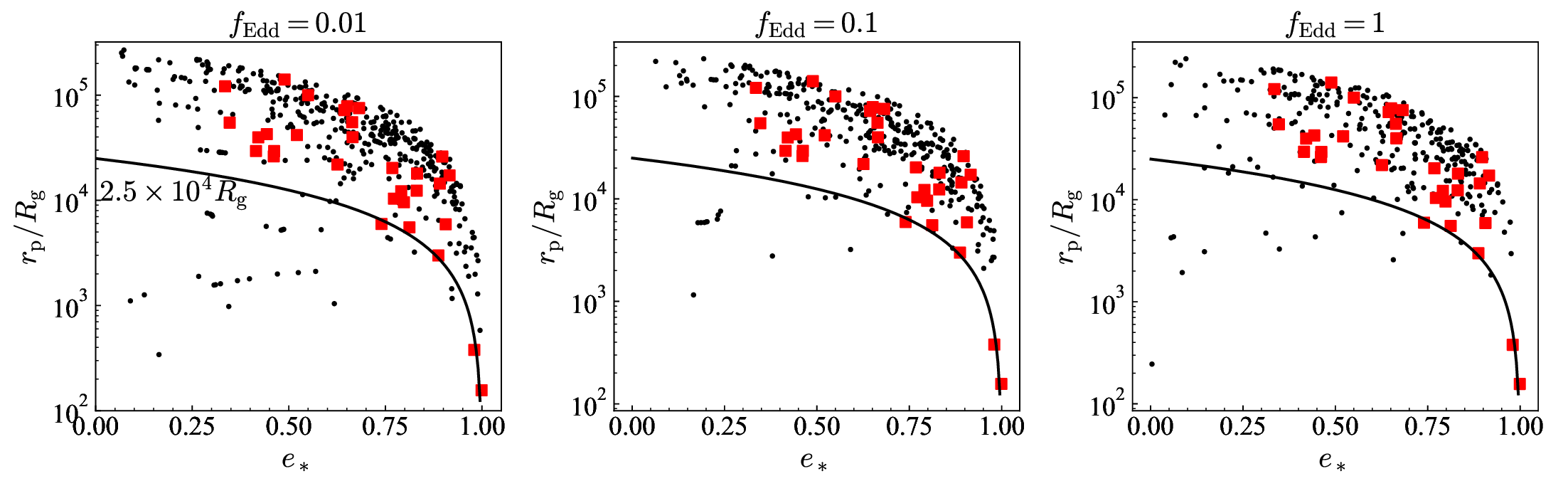}
\caption{The distribution of pericenter distances and eccentricities. The black dots show the results of the simulation and the red squares indicate the observations of S-stars. The black solid line depicts the semimajor axis of $2.5 \times 10^4 \, R_{\rm g}$ ($\sim$ 1000 au). The three panels from left to right represent the cases of $f_{\rm Edd}=0.01$, $f_{\rm Edd}=0.1$, and $f_{\rm Edd}=1$, respectively.}
\label{fig:a_distribution}
\end{figure*}

To compare our model predictions with observed kinematics of S-stars, we simulate the evolution of 500 stars with different initial orbital parameters. $m_{*}=10 \, {M}_{\odot}$ and $r_{*}=5 \, { R}_{\odot}$ are adopted in the calculations, which is the typical value for S-stars in the Galactic center \citep[e.g.,][]{Habibi2017}. The following initial conditions of the stars are assumed: (1) the initial inclination distribution of the stars is isotropic; (2) the argument of periapses is uniformly selected from 0 to $2\pi$; (3) the initial distribution of eccentricities follows a thermal eccentricity distribution; (4) the initial semimajor axis is in the range of $10^2-\,3\times 10^5 \, R_{\rm g}$, with a spatial number density distribution $n(r) \propto r^{-1.3}$. Due to the short migration timescale for stars embedded in the disk, we postulate that the star has either fallen into the SMBH or been tidally disrupted when the star has been fully immersed into the disk or when its pericenter distance is less than the tidal disruption radius ($ 44 \,R_{\rm g}$ for 10 ${M}_{\odot}$ star). Under different SMBH accretion rates, the distributions of stellar orbits after the evolution of stars for 5 Myr are shown in Figure \ref{fig:a_distribution}, where $r_{\rm p}=a_*(1-e_*)$ is the pericenter distance. 413, 335, and 294 stars still survive in the simulation (indicated by black dots), for $f_{\rm Edd} = 0.01$, $f_{\rm Edd} = 0.1$, and $f_{\rm Edd} = 1$, respectively. The orbital parameters of the observed S-stars are selected from \citet{Ali2020}, which is marked with red squares (deleting seven sources belong to the CWD). We can see that the positions of S-stars stay above the black line, which correspond to the semimajor axis of $2.5 \times 10^4 \, R_{\rm g}$ ($\sim$ 1000 au). Our simulations with $f_{\rm Edd} \gtrsim 0.1$ can roughly reproduce a similar distribution. The effect of drag force leads to a ``forbidden region," where the stars are difficult to survive due to the interaction with the accretion disk.

\begin{figure*}
\centering
\includegraphics[scale=0.47]{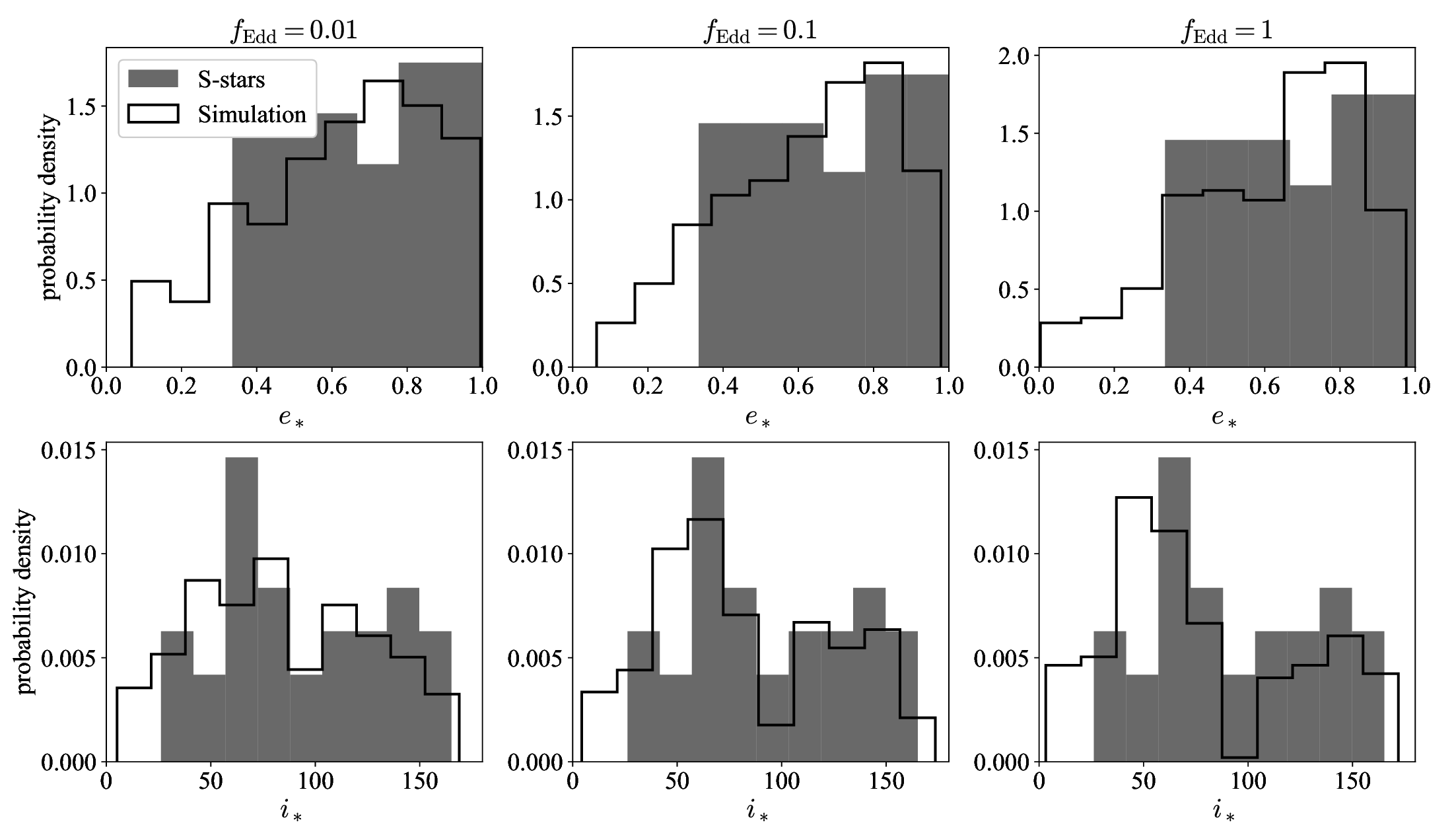} 
\caption{The observational and simulated distributions of eccentricities and inclinations of stars.}
\label{fig:kozai_distribution}
\end{figure*}

The comparison of the observational and the simulated distributions of eccentricities and inclinations is shown in Figure \ref{fig:kozai_distribution}, where the observational inclination of S-stars is relative to the plane of the CWD \citep{Paumard2006}. It can be found that most of the stars remain in high–eccentricity orbits after 5 Myr of evolution when assuming an initial thermal eccentricity distribution, which is more or less consistent with the observations. Meanwhile, our simulation results can also roughly reproduce the inclination distribution of S-stars, which has two peaks at $\sim 70^{\circ}$ and $ \sim 130^{\circ}$. For higher accretion rates, the inclination distribution will peak at a smaller angle because the effect of drag is more effective at higher surface densities of the accretion disk.

\section{Conclusions and Discussions}\label{sec:4}
The proximity of our Galactic center provides a unique opportunity to study the nuclear environments surrounding the SMBH with a spatial resolution much higher than that achievable to any other galaxy. Over the last three decades, precise measurements of the Galactic center reveal the unexpected presence of the young stars close to Sgr A*, which challenges the traditional star formation theories. Fortunately, the kinematic properties of these stars give us an opportunity to investigate their origin and orbital evolution. Based on the kinematics of the S-stars, \citet{Ali2020} argued that the distribution of inclinations deviates significantly from an isotropic distribution and speculated that they are arranged into two almost orthogonal disks. In this work, we explore the possible influences of an accretion disk on the dynamics of stars by assuming the accretion disk that once existed around Sgr A*. Stars could have formed in the outer gravitational unstable region of the accretion disk (e.g., stars in CWD). Considering the interaction of a spherical star cluster with this accretion disk, we roughly reproduce several observational features of S-stars. We find that the drag force exerted on stars during the collision with the disk may account for the absence of stars within $2.5 \times 10^4 \, R_{\rm g}$ ($\sim$ 1000 au) from Sgr A*, and the KL oscillations of stars due to the gravitational perturbation of the disk may be responsible for the two peaks of the distribution of S-star inclinations. At the same time, KL oscillations would also drive a majority of stars into high–eccentricity orbits.

The distributions of orbital parameters in our simulations are related to the timescale of star–disk interaction (namely, the duration of the past activity of Sgr A*) and the BH accretion rate, which are not well constrained. To explain the observational features of Fermi Bubbles, \citet{Zubovas2012} proposed Sgr A* was active with a mildly super-Eddington accretion rate 6 Myr ago and duration of the activity being about 1 Myr based on the ``quasar outflow" model. According to the jet model, \citet{Guo2012} suggested that the jet activity started about 1–3 Myr ago, with a duration of 0.1-0.5 Myr, and the accretion rate of $f_{\rm Edd} \sim 0.3$. However, \citet{Mou2014} argued that the Fermi Bubbles could also be inflated by winds launched from the hot accretion flow, which requires the active phase to last for about 10 Myr with $f_{\rm Edd} \sim 0.02$. In our results, the interaction timescale requires to be at least 3 Myr for the formation of a bimodal distribution of inclinations. However, 3 Myr is not enough for the effect of drag to generate a ``forbidden region" for stars within $\sim 1000$ au, in order to reproduce the distribution in Figure \ref{fig:a_distribution}, we estimate that the interaction timescale is at least $5$ Myr for the case of accretion rate $f_{\rm Edd} \gtrsim 0.1$. Our results are more or less similar to that derived from simulation of Fermi bubble, where the AGN activity may last several million years. It is worth noting that the results of the bimodal distribution of the inclination are insensitive to the distribution of the initial eccentricities. We test the case that the initial eccentricity ranges from 0 to 0.4; the bimodal distribution of inclinations can still form even though only approximately half of the stars are pumped into high eccentricity orbits.

In addition to the effects taken into account in this paper, there are several additional effects that may also affect the orbits of stars in the Galactic center. In our work, we assume a spherical distribution of stars, but the fluctuating stochastic anisotropy of a stellar bulge will generate a strong net gravitational torque on stellar orbits, which could randomize the direction of the stellar orbits via a process called vector resonant relaxation \citep{Rauch1996}. However, \citet{Panamarev2022} found that even 10 Myr is not long enough for vector resonant relaxation to randomize the orbital inclinations. \citet{Levin2022} proposed that if the rotational direction of a stellar disk is misaligned with that of its host nuclear star cluster, a process called resonant friction will disrupt the outer stellar disk, while the inner stellar disk will nearly remain intact and gradually align with the rotation of the star cluster. He estimated the case of the Galactic center and found that the effect of resonant friction becomes significant after 5 Myr \citep[see the case rotation 0.3 of Fig. 1 in][]{Levin2022}. It should be noted that the anisotropic distribution of stellar distribution in the Galactic center will lead to different results, which is worthy in our future investigation.

Several works proposed that the star will grow its mass through Bondi–Hoyle–Littleton accretion during crossing through the AGN disk \citep[e.g.,][]{davies2020,generozov2023}. If this is correct, accretion of gas onto the stars could lead to rejuvenation of the star, exhibiting characteristics of a young star \citep[e.g.][] {jermyn2021,cantiello2021,dittmann2021}, which may be the physical reason for the young age of S-stars. However, the accretion process during the star crossing the accretion disk still suffers many uncertainties. \citet{Lu2023} suggested that a strong bow shock generated by the collision may strip off the outer layers of the stars due to Roche-lobe overflow during the stars supersonically cross the accretion disk. We estimate and find that the Roche-lobe overflow will only occur for a crossing position that is less than $\sim 100\, R_{\rm g}$; therefore, the stripping of the outer layers of stars should not be important for S-stars.

~\\
\noindent We thank the referee for constructive comments and suggestions that help us to improve this paper. The work is supported by the National SKA Program of China (2022SKA0120101), the National Natural Science Foundation of China (grants U1931203 and  12233007) and the science research grants from the China Manned Space Project (No. CMS-CSST-2021-A06)

\bibliography{S-stars}{}
\bibliographystyle{aasjournal}

\end{document}